\documentclass[preprint,showpacs,preprintnumbers,amsmath,amssymb]{revtex4}
\pdfoutput=1
\usepackage{dcolumn}
\usepackage{graphicx}
\usepackage{graphicx,amssymb,amsmath}
\usepackage{graphics}
\usepackage{ulem}
\usepackage{color}
\def\r{\boldsymbol{r}}
\def\x{\boldsymbol{x}}

\def\kt{k_{\rm B} T}

\def\beq{\begin{equation}}
\def\eeq{\end{equation}}
\def\bea{\begin{eqnarray}}
\def\eea{\end{eqnarray}}

\begin{document}

\title{Electrostatic interactions in concentrated protein solutions}
\author{Shradha Mishra\footnote{{shradha@phys.ksu.edu, Present address: Department of physics, Indian Institute of technology Kharagpur,
Kharagpur India, 721302 }} and Jeremy D.\ Schmit\footnote{schmit@phys.ksu.edu}}
\affiliation{Department of Physics, Kansas State University,
 Manhattan, KS 66506, USA}

\begin{abstract}
We present an approximate method for calculating the electrostatic
free energy of concentrated protein solutions.  Our method uses a
cell model and accounts for both the coulomb energy and the entropic
cost of Donnan salt partitioning. The former term is calculated by
linearizing the Poisson-Boltzmann equation around a nonzero average
potential, while the second term is calculated using a jellium
approximation that is empirically modified to reproduce the dilute
solution limit. When combined with a short-ranged binding
interaction, calculated using the mean spherical approximation, our
model reproduces osmotic pressure measurements of bovine serum
albumin solutions.  We also use our free energy to calculate the
salt-dependent shift in the critical temperature of lysozyme
solutions and show why the predicted salt partitioning between the
dilute and dense phases has proven experimentally elusive.
\end{abstract}

\maketitle

\section{Introduction.}
Dense protein solutions are encountered in the cytoplasm and {\em in
vitro} situations like pharmaceutical formulations, crystallization
screens, and ultrafiltration
\cite{Zimmerman1993,Shire2004,McPherson2004}.  In all these cases,
the stability of the solution depends on sufficient electrostatic
repulsion to overcome the short-range attraction of H-bond,
hydrophobic, and van der Waals interactions.  Electrostatic
interactions are easily adjusted through changes in the pH or salt
concentration providing a convenient experimental means to
manipulate the phase behavior. However, they are difficult to model
theoretically due to the long range nature of the coulomb force and
the nonlinearity of salt screening. Because of this, most
theoretical work on protein-protein interactions has focused on
dilute solution properties where the electrostatic free energy is
dominated by the coulomb energy and can be treated using two-body
potentials \cite{Tavares:2004b,Elcock2001,Coen1996}.  We have
recently shown that the electrostatic free energy of protein
association is dominated by the change in salt ion entropy, which
renders the net interaction strongly non-pairwise
\cite{Schmit2011gel}.  In this paper we extend these results in
order to model electrostatic effects in non-ideal protein solutions.

We test our theory against pH dependent measurements of the osmotic
pressure and salt effects on the liquid-liquid phase separation of
protein solutions.  The osmotic pressure provides a direct test of
the effective interparticle repulsion, while the latter phenomenon
has attracted considerable attention due to the finding that
fluctuations associated with the liquid-liquid critical point have
been shown to accelerate crystal nucleation \cite{tenWolde1997}.
Liquid-liquid separation is analogous to the liquid-vapor
coexistence in small molecules except that in proteins the
liquid-liquid binodal is found entirely below the solid (crystal)
solubility line due to the short range of the protein-protein
attractive forces \cite{Hagen1994,Asherie:1996}.  Previous work has
primarily focused on temperature as a means of controlling a phase
behavior. However, temperature is a poor variable in protein systems
because the accessible temperature range is limited to $\sim 20\%$
due to the freezing point of water and the thermal denaturation of
the proteins.  Our work shows how pH and salt can be used to
manipulate the phase boundary into the accessible range.

\section{Model}
Our calculations are based on the following free energy of the
ternary protein-salt-solvent system
\begin{equation}
f(\eta)=F(\eta)/N=f_{\rm hc}(\eta)+f_{\rm att}(\eta)+f_{\rm
salt}(\eta)+f_{\rm coulomb}(\eta) \label{eq:Ftotal}
\end{equation}
where $f(\eta)$ is the total free energy per protein as function of 
protein concentration $\eta$. 
$f_{\rm hc}$ is the mixing entropy of spherical proteins interacting
by a hard core potential, $f_{\rm att}$ is the free energy due to
short range attractions between the proteins, $f_{\rm salt}$ is the
mixing entropy of the salt ions, and $f_{\rm coulomb}$ is the
coulomb energy of the system.  These terms are explicit functions of
$\eta=N \sigma^3/6V$, the volume fraction occupied by the proteins,
where $\sigma$ is the protein diameter and $N/V$ is protein number
density.  We discuss each of these terms in detail below.

\subsection{Free energy of attractive spheres}
The terms $f_{\rm hc}$ and $f_{\rm att}$ represent the free energy
of a solution of attractive spheres.  We adopt the attractive Yukawa
potential to describe the short range interaction
\begin{equation}
U(r) = \left\{\begin{array}{ll}
\infty & r \le \sigma;\\
-\epsilon(T) \frac{\exp[-z(r-\sigma)]\sigma}{r} & r > \sigma
\end{array}\right. \label{yukawa}
\end{equation}
where $z$ is a parameter describing the range of interaction,
$\sigma$ is the hard sphere diameter, and $\epsilon$ is the
temperature dependent strength of interaction.  Following previous
work \cite{Gogelein2008,Fu2003}, we take $z=4$ reflecting the short
range nature of the hydrophobic and H-bond interactions that
dominate protein-protein attractions.

Eq. \ref{yukawa} gives the binding energy of a two-body
protein-protein interaction as a function of the center-to-center
distance $r$.  In order to derive macroscopic properties of the
protein solution, we need to know the average binding energy per
particle as a function of the protein concentration.  We obtain this
using the Mean Spherical Approximation
\cite{Tang2003,Durand-Vidal2000}
the mean spherical approximation (MSA) is the method used to
obtain the analytical solution of the radial 
distribution function of particles. The thermodynamic properties obtained
from the MSA are in good agreement with results obtained from the
Monte Carlo simulation. 

\begin{figure}[ht]
\vspace{0.6 cm}
\begin{center}
\includegraphics[width=0.7\linewidth]{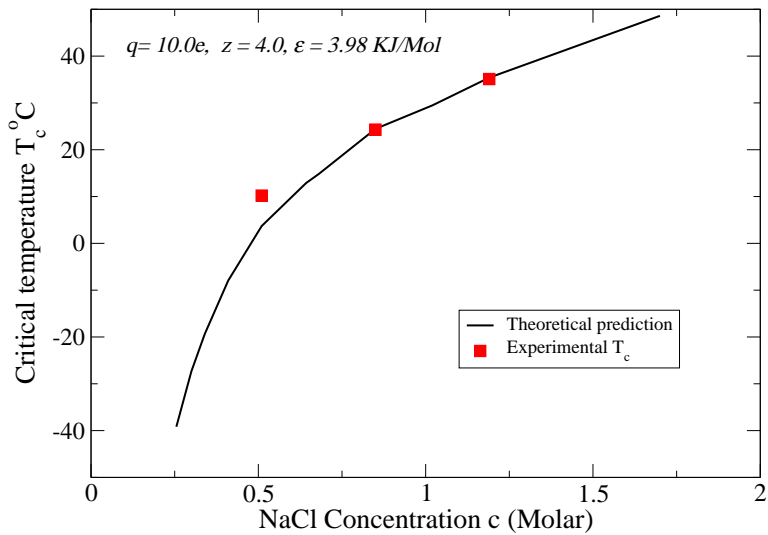}
\end{center}
\caption{Plot of critical temperature $T_c$ vs. salt concentration
in Molar for protein charge q = 10.0} \label{fig:temp_c}
\end{figure}

Within this approximation the binding energy density is \cite{Duh1997,Fu2003}
\begin{equation}
f_{\rm att}=\frac{\alpha_0}{\Phi_0} \beta \epsilon - \frac{z^3}{6
\eta} \bigg[\mathcal{F}(X) - \mathcal{F}(Y) - (X - Y) \frac{d
\mathcal{F}(Y)}{d Y}\bigg] \label{eq:msa}
\end{equation}
where $X$ and $Y$ are the variable defined in \ref{eqA28} and \ref{eqA29}.
 $z$ is the parameter for range of interaction in the Yukawa
potential and $\eta$ is the volume fraction of the proteins.
$\alpha_0$, $\Phi_0$, $\mathcal{F}(X)$ and its first derivative are
defined in Eqs. \ref{eqA23}, \ref{eqA25}, \ref{eqA31} and
\ref{eqA32}.  This binding energy is partially offset by the loss of
translational entropy at the high concentration.  This contribution
is given by the Carnahan-Starling expression \cite{Carnahan1969}
\begin{equation}
f_{\rm hc}(\eta)/\kt=\eta^2\frac{4-3\eta}{(1-\eta)^2}. \label{eq:cs}
\end{equation}
Advantage of using MSA for our case is that, we get an expression
for free energy density in terms of density or 
volume fraction $\eta$ of protein.
We can compute equation of state, osmotic pressure and many other 
thermodynamic quatities using this free energy density. 
MSA is based on the inverse temperature expansion of the free energy, 
hence it gives better result for higher temperatures. In  
our phase coexistence curve \ref{fig:coexx} and \ref{fig:temp_c}
 we find our theoretical curves
are more promising for large salt or when we have larger critical. 
temperautre. 
\subsection{Electrostatic terms}
To solve for the electrostatic free energy, we adopt a cell model
\cite{Wall1957} in which each protein is surrounded by a spherical
shell of solvent of radius $b$.  The thickness of this solvent layer
is chosen to reproduce the volume fraction occupied by the protein
$\eta=(\sigma/2b)^3$. We assume a protein solution in osmotic
equilibrium with a reservoir of symmetric, monovalent salt of
concentration $c_s$.  The proteins carry a charge $qe_p$, where
$e_p$ is the proton charge, that we take to be uniformly distributed
over their spherical surface. The protein charge will perturb the
salt ion concentration resulting in a local enrichment of
counterions and a local depletion of coions. The extent of this
ionic perturbation is a competition between the coulomb interaction
of the salt ions with the electrostatic potential $\Phi$ and the
entropic cost of enriching/depleting the counterion/coion
populations.  These considerations are reflected in the
electrostatic free energy $f_{\rm ES}=f_{\rm coulomb}+f_{\rm salt}$
where
\begin{eqnarray}
f_{\rm coulomb}& =& \frac{\varepsilon}{2} \int_V{(\nabla \Psi)^2 d^3 \r} \label{eq:coulomb}\\
f_{\rm salt}/k_BT &=&\int_{V'} [c_+\ln(c_+/c_s)-c_+ + c_s]+
[c_-\ln(c_-/c_s)-c_- + c_s] d^3\r, \label{eq:ionentropy}
\end{eqnarray}
where $c_\pm$ are the local cation/anion concentrations, $V$ is the
cell volume, $V'$ is the solvent accessible cell volume, and
$\varepsilon$ is the local permeability which we take to be
$80\varepsilon_0$. The electrostatic potential can be expressed in
terms of the charge distribution using the Poisson equation
\begin{equation}
-\nabla\cdot(\varepsilon \nabla \Psi(\r))=\rho_p(\r)+e_p(c_+(\r) -
c_-(\r)) \label{eq:poisson}
\end{equation}
where $\rho_p$ is the charge density of the protein.  The ion
concentrations can be found by minimizing $f_{\rm ES}$ with respect
to $c_\pm$ after integrating Eq. \ref{eq:coulomb} by parts and
applying Eq. \ref{eq:poisson}.  This resulting concentrations are
\begin{equation}
c_\pm(\r)=c_s e^{\mp e_p\Psi(\r)/k_BT}, \label{eq:ionboltzmann}
\end{equation}
which, with Eq. \ref{eq:poisson}, gives the well known
Poisson-Boltzmann equation
\begin{equation}
-\nabla\cdot(\varepsilon \nabla \Psi(\r))=-\rho_p(\r)-e_p c_s(
e^{-e_p \Psi(\r)/k_BT}- e^{e_p \Psi(\r)/k_BT}). \label{eq:PB}
\end{equation}
Our strategy is to develop approximate solutions of Eq. \ref{eq:PB}
that can be used in Eqs. \ref{eq:coulomb} and \ref{eq:ionentropy} to
calculate the free energy as a function of protein concentration,
charge, and salt concentration.

Within the cell geometry, the Poisson-Boltzmann equation has the
approximate solution
\begin{eqnarray}
\Psi(x)&=&\phi(x)+\phi_0 \nonumber \\
&=&C\left(\frac{e^{\beta-x}(\beta-1)}{x}+\frac{e^{x-\beta}(\beta+1)}{x}\right)
-\tanh \phi_0 + \phi_0, \label{eq:shiftedDHpot}
\end{eqnarray}
where $\alpha = (\cosh \bar{\phi})^{1/2}\kappa \sigma/2$, $\beta =
(\cosh \bar{\phi})^{1/2}\kappa b$ and $C$ is a constant defined in
the appendix. Eq. \ref{eq:shiftedDHpot} is derived by linearizing
the Poisson-Boltzmann equation around a reference potential $\phi_0$
($\phi_0=0$ in the Debye-Huckel solution).  The nonzero reference is
necessary to model concentrated solutions of charged proteins, where
the Donnan effect ensures that the potential never approaches zero.

Eq. \ref{eq:shiftedDHpot} can be used with Eq. \ref{eq:coulomb} to
determine the coulomb energy per particle.  A straightforward
integration yields (see appendix)
\begin{eqnarray}
f_{\rm
coulomb}&=&-2C^2\beta^3+\frac{C^2}{\alpha}((\beta-1)e^{-\alpha+\beta}+(\beta+1)e^{\alpha-\beta})^2
\nonumber \\
&&-\frac{C^2}{2}((\beta+1)^2e^{2(\alpha-\beta)}-(\beta-1)^2e^{-2(\alpha-\beta)}-4(\beta^2-1)\alpha)
\label{fcoulomb}
\end{eqnarray}
To compute $f_{\rm salt}$ we first combine Eqs. \ref{eq:ionentropy}
and \ref{eq:ionboltzmann} to obtain
\begin{equation}
f_{\rm salt}/k_BT =2 c_s \int_{V'} \bigg[\left(\frac{e_p
\phi}{k_BT}\right)\sinh \left(\frac{e_p \phi}{k_BT}\right) - \cosh
\left(\frac{e_p \phi}{k_BT}\right) +1\bigg] d^3\x
\label{eq:ionintegration}
\end{equation}
This expression does not lend itself to direct integration, however,
at high protein concentrations the potential varies weakly in the
voids between proteins and it is reasonable to replace it with an
average value $\phi(\x)\rightarrow \bar{\phi}$ (the jellium
approximation) so that $c_\pm$ are constant \cite{Levin2003}. Charge
neutrality requires
\begin{eqnarray}
q&=&-v_{\rm ion}(c_+-c_-) \label{eq:chargeconservation}\\
&=&2v_{\rm ion} c_s \sinh(e\bar{\phi}/k_BT) \\
\frac{e\bar{\phi}}{k_BT}&=&\sinh^{-1}\frac{q}{2 v_{\rm ion} c_s}
\label{eq:psibar},
\end{eqnarray}
where $v_{\rm ion}$, the solvent volume associated with each
protein, is given by
\begin{equation}
v_{\rm ion}=v_p\left(\frac{1}{\eta}-1\right), \label{eq:VwaterDense}
\end{equation}
where $v_p=\pi \sigma^3/6$ is the volume of a single protein.

\begin{figure}[ht]
\vspace{0.6 cm}
\begin{center}
\includegraphics[width=0.7\linewidth]{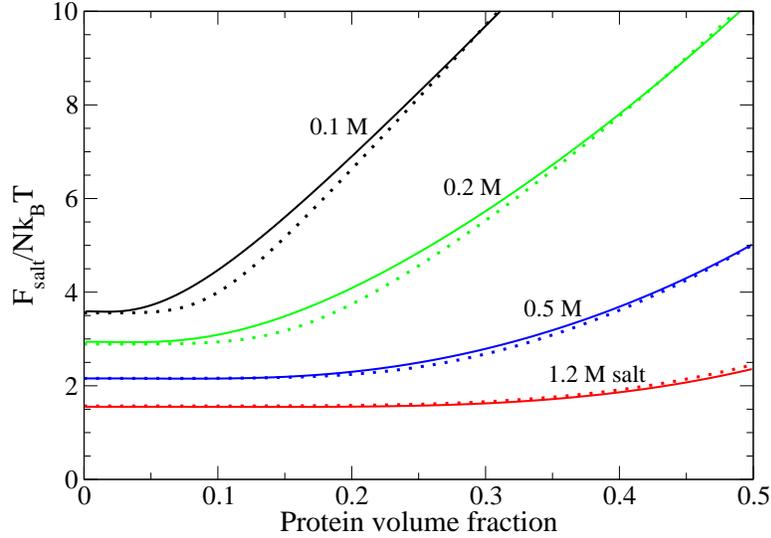}
\end{center}
\caption{Comparison of two methods for calculating the salt entropy
per particle.  Solid lines show a numerical integration of Eq.
\ref{eq:ionentropy} with the potential given by Eq.
\ref{eq:shiftedDHpot}.  Dashed lines show Eq.
\ref{eq:ionentropyjellium} with the screening layer volume given by
\ref{eq:Vwaterapprox}.  $\sigma/2=16$nm, $b=(\sigma/2+1.9/\kappa)$,
$q=10.3$. The dashed lines have been shifted by an unimportant
constant to facilitate comparison. \label{fig:jelliumapprox}}
\end{figure}

Combining Eqs. \ref{eq:ionentropy} and \ref{eq:psibar} we find an
expression for the salt entropy per protein at high protein
concentrations
\begin{equation}
f_{\rm salt}/Nk_BT=q(\sinh^{-1}\xi - \sqrt{1+\xi^{-1}}+\xi^{-1})
\label{eq:ionentropyjellium}
\end{equation}
where $\xi=q/2v_{\rm ion} c_s$.  This expression is an excellent
approximation for high concentration solutions, but fails in the
dilute limit where it erroneously predicts that the entropic cost of
the ion screening layers approaches zero.  The problem can be traced
to Eq. \ref{eq:VwaterDense} which implies that the ion screening
layer can become arbitrarily large.  In reality, in dilute solutions
the screening layer is confined to a shell with a thickness on the
order of a Debye length.  Because of this, the salt entropy will
saturate at a minimum value when the concentration drops below a
critical value $\eta_0$.  This behavior can be obtained from our
model by a numerical integration of Eq. \ref{eq:ionentropy} with the
potential given by \ref{eq:shiftedDHpot}.  However, a more
computationally efficient solution is to modify Eq.
\ref{eq:VwaterDense} to give the correct asymptotic behavior.  This
can be done with the following functional form
\begin{equation}
v_w^{\text{eff}}=v_p\left(\frac{1}{(\eta^n+\eta_0^n)^{1/n}}-1\right).
\label{eq:Vwaterapprox}
\end{equation}
Here $n$ and $\eta_0$ are adjustable parameters that give optimal
results when $n=5$ and $\eta_0=(1+3.8/\kappa \sigma)^3$ where
$\kappa^{-1}$ is the Debye screening length.  Fig.
\ref{fig:jelliumapprox} shows an excellent agreement in the salt
entropy calculated using these two methods. Therefore, the remainder
of the paper will utilize the more efficient effective water shell
calculation of salt entropy.

\section{Results and Discussion}
\subsection{Osmotic pressure}
\begin{figure}[ht]
\vspace{0.6 cm}
\begin{center}
\includegraphics[width=0.7\linewidth]{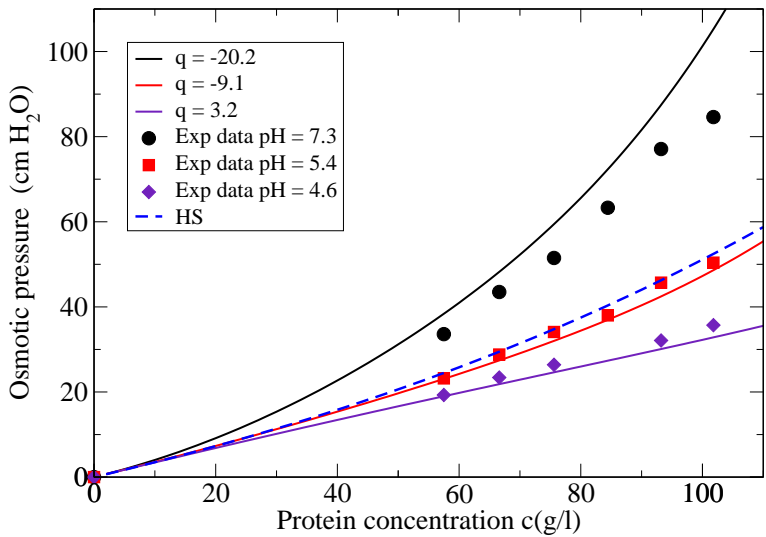}
\end{center}
\caption{Plot of osmotic pressure vs. protein concentration for
three different protein charges, $q = -20.2$, $q = -9.1$ and $q =
3.2$ (solid lines). Points are data from \cite{Kanal1994} for three
different pH values pH = $7.3$, $5.4$ and $4.6$. The measured
protein charges at these pH values are obtained from
\cite{Vilker1981}. $\epsilon = 2.98$ kJ/mol is our fitting
parameter.
Temperature and salt concentration is same  as in the experiment
\cite{Kanal1994}. \label{fig:OP_Z}}
\end{figure}

As an initial test of our theory we compute the osmotic pressure of
charged protein solutions
\begin{equation}
\Pi =  \eta \frac{\partial f}{\partial \eta}.
\end{equation}
Fig. \ref{fig:OP_Z}
compares the theory to experimental measurements of bovine serum
albumin \cite{Kanal1994}.  We limit our comparison to pH values
where the protein charge has been measured \cite{Vilker1981},
however the resulting charge values $q=-20.2$ (pH = 7.3), $q=-9.1$
(pH = 5.4), and $q=3.2$ (pH = 4.6) cover a sufficiently large range
to provide a rigorous test of the model.  We use a protein radius of
2.4 nm \cite{Minton1982} and the single fitting parameter is
$\epsilon=1.22 k_B T = 2.98$ kJ/mol.  The agreement is generally
good, with an average error of 6.5\% for $q = 3.2$ and $-9.1$ and a
larger 14\% error (12\% excluding the outlier at 100 g/l) at $q =
-20.2$.

In earlier work these data were modeled by fitting an effective
protein volume for each pH \cite{Minton1995}. These effective
volumes ranged from slightly {\em negative} near the isoelectric
point to four-fold greater than the actual volume when the protein
carries a charge of $\sim 60.0$ \cite{Salis2011}. Our modeling
suggests that these trends can be explained by nonspecific protein
binding competing with electrostatic monopole repulsion (mediated by
the salt ions).

\begin{figure}[ht]
\vspace{0.6 cm}
\begin{center}
\includegraphics[width=0.4\linewidth]{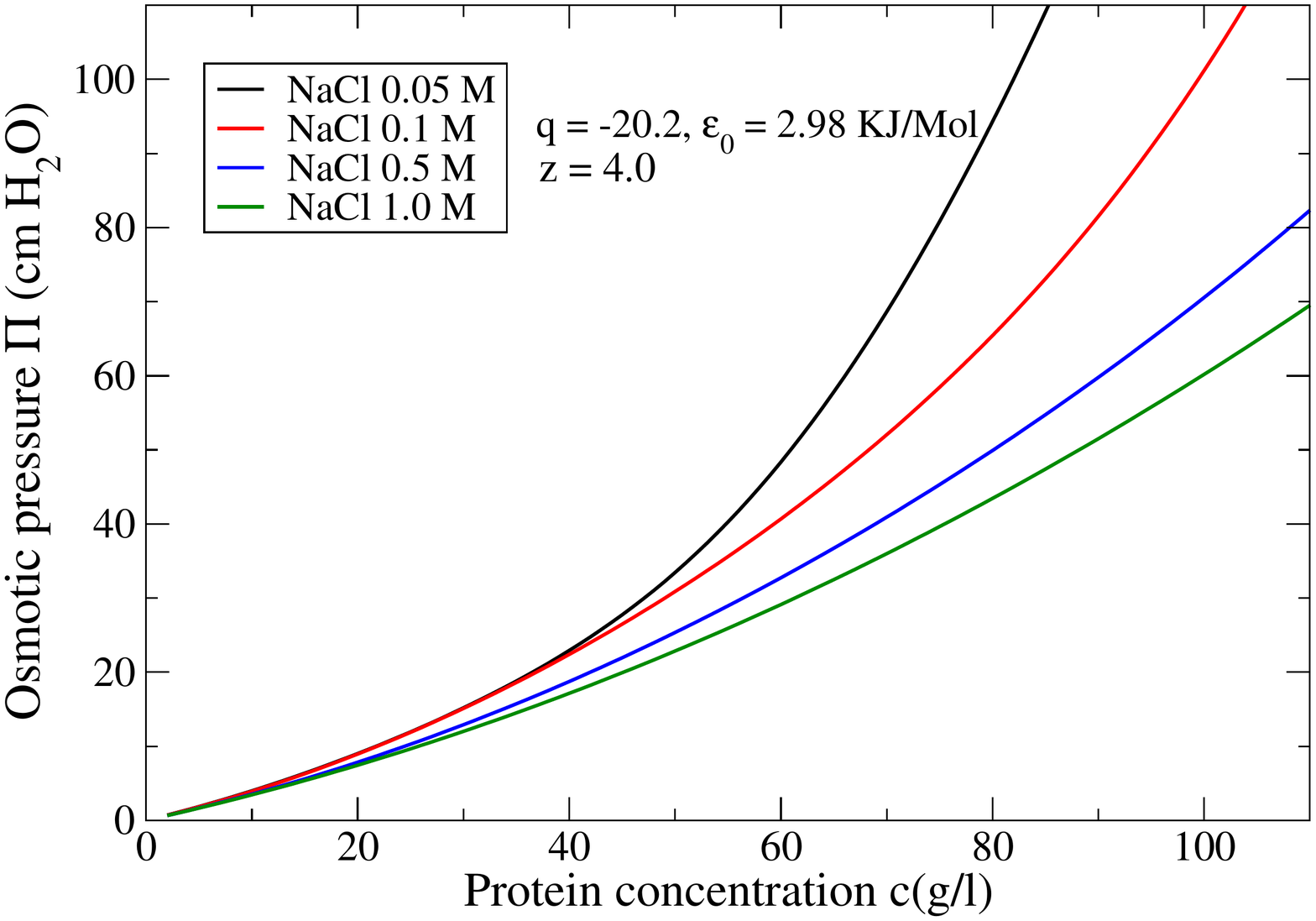}
\includegraphics[width=0.4\linewidth]{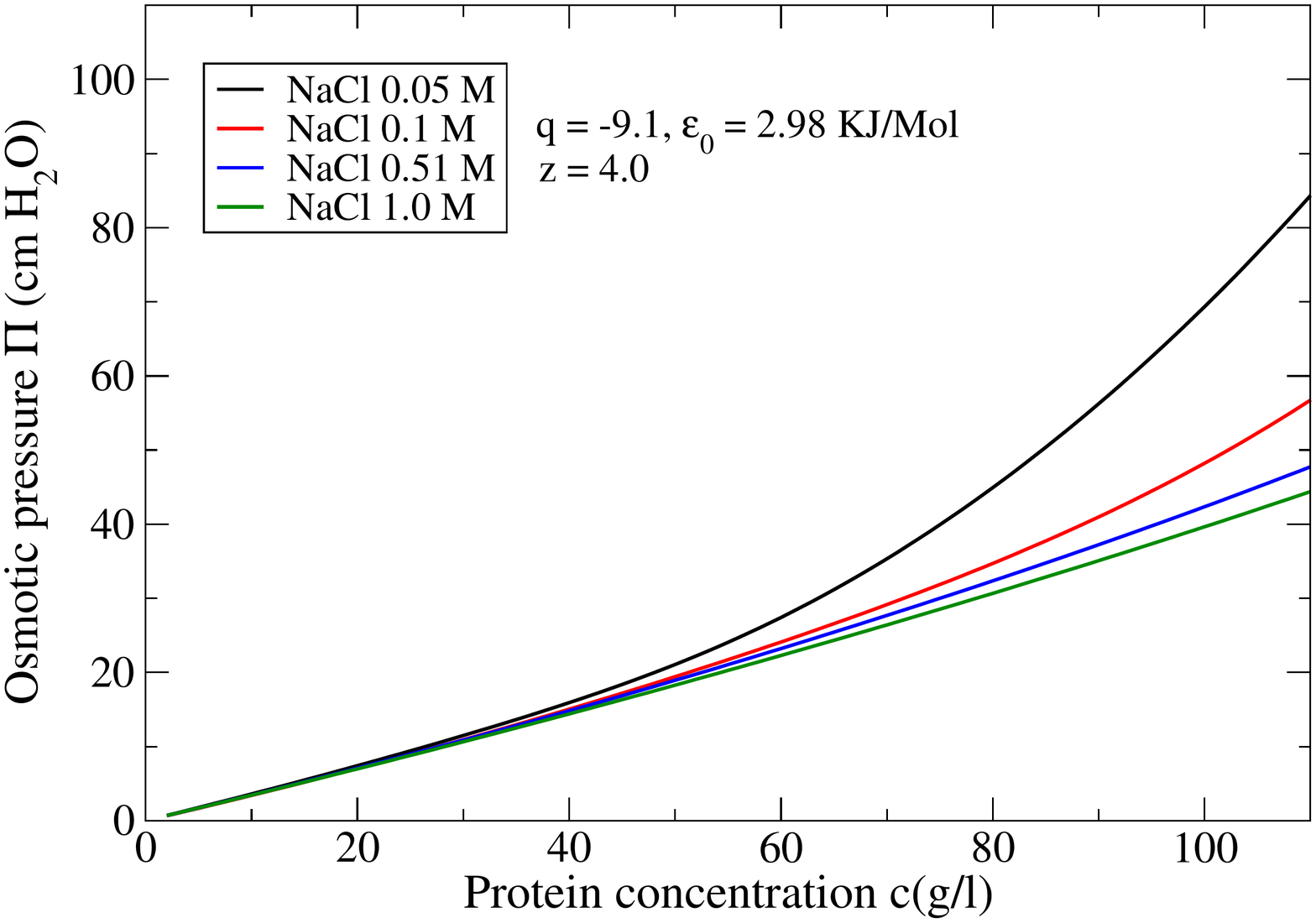}
\end{center}
\caption{Plot of osmotic pressure vs. protein concentration for salt
concentrations from 0.05M to 1.0M. The two panels are for  protein
charges $q = -20.2$ (left) and $q = -9.1$ (right).}
\label{fig:OP_Salt}
\end{figure}

Fig. \ref{fig:OP_Salt} shows predictions for the osmotic pressure as
a function of salt for two values of the protein charge.  Not
surprisingly, greater salt concentrations result in lower osmotic
pressures.  This is explained by the fact that at higher salt
concentrations the neutralizing counterions lie closer to the
surface of the protein.  Therefore, neighboring proteins will impose
fewer constraints on the entropy of the screening layers.

\begin{figure}[ht]
\vspace{0.6 cm}
\begin{center}
\includegraphics[width=0.7\linewidth]{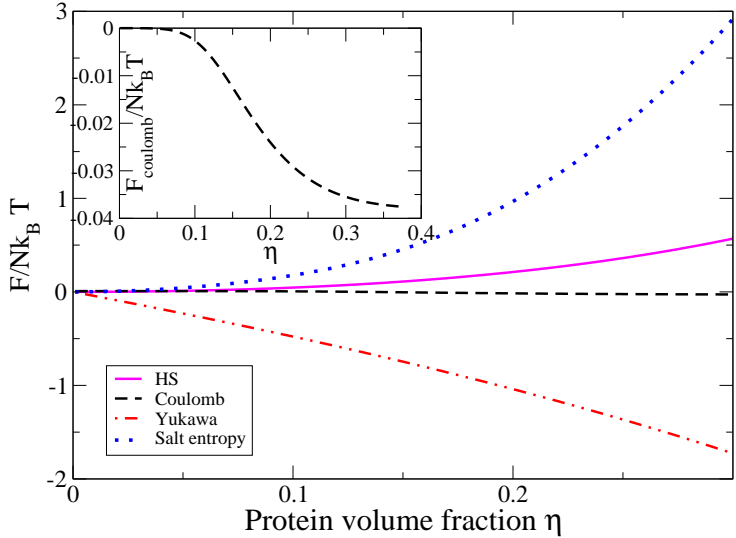}
\end{center}
\caption{Plot of the contributions to the free energy $F/NK_BT$. The
four curves are $f_{\rm hc}$ (solid), $f_{\rm coulomb}$ (dashes),
$f_{\rm att}$ (dot-dash), $f_{\rm salt}$ (dots).  The coulomb term
is very small (inset) compared to other three terms.} \label{fig:Fterms}
\end{figure}

Fig. \ref{fig:Fterms} shows the component terms of the free energy
as a function of the concentration.  It is immediately apparent that
the coulomb energy is minuscule in comparison with the other terms.
This observation is somewhat misleading because the total free
energy is a small residual from the addition of large terms with
opposing signs.  Therefore, the coulomb energy perturbation does
give a significant quantitative improvement. On the other hand, if
quantitative results are not required, a good estimate of pH and
salt effects can be obtained from Eq. \ref{eq:ionentropyjellium}
while neglecting the more difficult to calculate coulomb energy.

\begin{figure}[ht]
\vspace{0.6 cm}
\begin{center}
\includegraphics[width=0.7\linewidth]{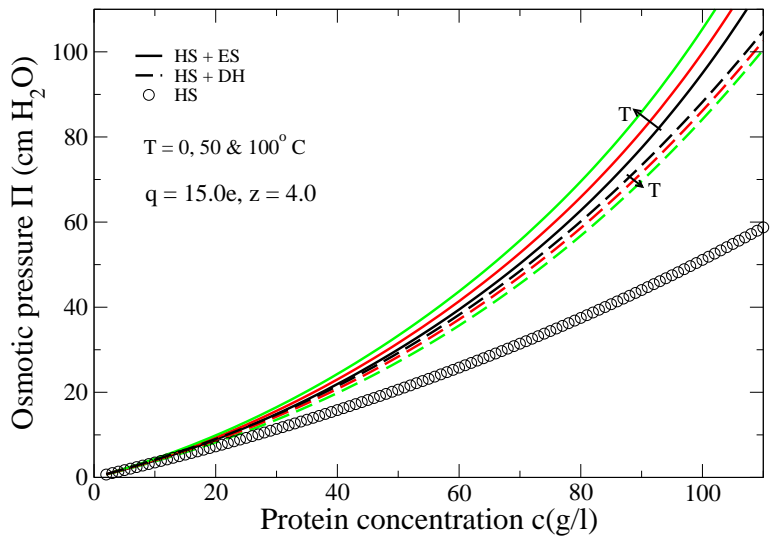}
\end{center}
\caption{Plot of osmotic pressure (OP) $\Pi$ vs. protein concetration $\eta$
 for two
different representations of the electrostatic interaction. Solid
curves show the OP calculated for hard spheres with the entropy
dominated interaction given by Eqs. \ref{fcoulomb} and
\ref{eq:ionentropyjellium}. The three curves show the expected
increase in OP with temperature.  The dashed curves show the OP
calculated for hard spheres with a Debye-Huckel type electrostatic
interaction. This model shows a non-physical decrease in the OP with
temperature.} \label{fig:DHvDonnan}
\end{figure}

Another observation from Fig. \ref{fig:Fterms} is that the repulsive
terms are entropic in origin while the attractive terms are
energetic.  This has important consequences from the temperature
dependence of the pressure.  In Fig. \ref{fig:DHvDonnan} we plot the
temperature dependence of the osmotic pressure for two systems of
hard spheres.  The first system interacts by the entropy dominated
electrostatic interaction (Eqs. \ref{fcoulomb} and
\ref{eq:ionentropyjellium}) while the second interacts by a
repulsive Yukawa (Debye-Huckel) potential that is energy dominated
(calculated using the MSA). We see that the temperature has
qualitatively different effects on the two systems. In the Yukawa
system the osmotic pressure actually decreases with temperature
because the reduced Boltzmann weight given to the repulsive
interaction at higher temperatures effectively frees volume for the
spheres to explore. It is only with the entropy dominated
electrostatic interaction that we recover the physically reasonable
result that the osmotic pressure should increase with temperature
\cite{Piazza:1999}.

\subsection{Liquid-liquid phase separation}

The solution free energy Eq. \ref{eq:Ftotal} contains a binding
energy term that favors dense states competing with entropic terms
that favor dilute states.  This means that under some conditions the
solution may phase separate in order to maximize the free energy
contribution from these extreme states.  The conditions for an
equilibrium between two phases of unequal densities are
\begin{eqnarray}
\mu'_p&=&\mu''_p \\
\mu'_+&=&\mu''_+ \label{eq:MuCation}\\
\mu'_-&=&\mu''_- \label{eq:MuAnion}
\end{eqnarray}
where $\mu'$ and $\mu''$ are the chemical potential of the dilute
and dense phases, respectively, and the subscripts represent
proteins, cations, and anions.  The conditions for the salt ions are
satisfied by Eq. \ref{eq:ionboltzmann} which captures both mixing
entropy and coulomb energy contributions to the chemical potential.
Eqs. \ref{eq:MuCation} and \ref{eq:MuAnion} are, therefore, built
into our free energy.  Thus, the slope $\partial F/\partial \eta$ is
an effective chemical potential for the protein that includes the
effects of maintaining the salt equilibrium.

\begin{figure}[ht]
\vspace{0.6 cm}
\begin{center}
\includegraphics[width=0.7\linewidth]{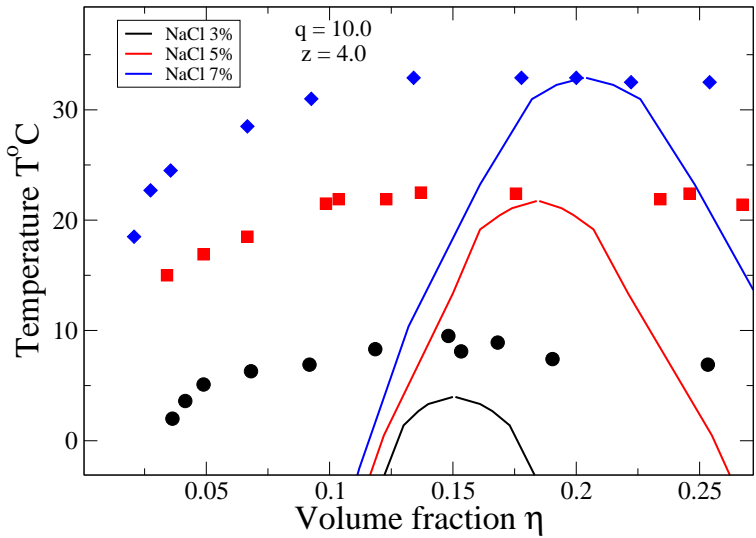}
\end{center}
\caption{Plot of liquid-liquid coexistence curve for three different
salt concentrations NaCl $3\%$, $5\%$ and $7\%$. Data points are
experimental curve from \cite{Muschol1995}. We have good predictions
for critical temperature. Other parameters protein charge $q =
10.0$, range of Yukawa interaction $z= 4.0$ and binding energy
$\epsilon = 3.98$ kJ/mol. We kept binding energy as our fitting
parameter and fixed all other parameters.} \label{fig:coexx}
\end{figure}

The densities of coexisting phases are found by numerically
searching for a line with two points tangent to $F(\eta)$
\cite{Chaikin2000}. In Fig. \ref{fig:coexx} we plot the coexistence
curve predicted by Eq. \ref{eq:Ftotal} for three different NaCl
concentrations ($ 3\%$, $ 5\%$ and $ 7\%$ w/v) for a particle of
diameter $\sigma=3.2$nm and charge $q=10.3e$. These parameters
correspond to the pH $4.5$ conditions used by Muschol and
Rosenberger \cite{Muschol:1997} to determine the liquid-liquid
binodal curves for lysozyme.  The single free parameter,
$\epsilon=3.98$ kJ/mol, has been set to reproduce the critical
temperature under 7\% salt.

It is immediately obvious that the theory fails to capture the width
of the lysozyme coexistence curves.  This is because proteins have
much broader binodal curves than systems of smooth spheres
\cite{Lomakin1996}. The reasons for this are not clear, but possible
factors include anisotropic or directional binding
\cite{Kern:2003,Bianchi:2008,Liu:2009}, temperature dependence of
the short range attraction \cite{Gogelein2008}, asphericity
\cite{Wang2011a}, and the range of the attraction \cite{Fu2003}.
However, the theory does a reasonable job of reproducing the
salt-dependent shift in the critical temperature, although as we
observed with the osmotic pressure, the theory somewhat
over-predicts the trend under the most extreme conditions (low salt
or high charge).

A key feature of this theory is the partitioning of ions between the
dilute and dense phases by the Donnan effect.  However, experiments
to observe this partitioning have yielded negative results
\cite{Muschol:1997}. Using Eq. \ref{eq:chargeconservation}
 we can see why the partitioning was experimentally elusive.
At pH 4.5 lysozyme has a calculated charge of approximately $Z=+10$.
If we take the protein volume to be 17 nm$^3$ and assume that the
dense phase has a density $\eta=1/3$, then for 0.5M salt conditions
Eq. \ref{eq:psibar} gives a potential of $e\phi/k_BT=0.46$.  From
Eq. \ref{eq:ionboltzmann} we get a counterion concentration of
$1.58\ c_0$ within the solvent fraction of the dense phase.  To get
the apparent concentration within the total phase we multiply by
$1-\eta$ to account for the excluded volume of the protein.  We find
that the apparent counterion concentration in the dense phase is
nearly unchanged from the bulk value.  However, this cancelation is
entirely a coincidence of the conditions used in Ref
\cite{Muschol:1997}. A similar calculation for the coions yields an
apparent concentration of $0.4 c_0$, however, this may have been
missed due to difficulties associated with the sodium assay [M.
Muschol personal communication].

\section{Conclusion}
We have presented an approximate method for calculating
electrostatic effects in dense protein solutions.  Our method
captures the non-pairwise nature of electrostatic interactions at
high concentrations, yet requires only a single particle calculation
within a cell model.  Such methods may enable the rational
manipulation of protein solution behavior and phase diagrams using
pH and salt concentration as control parameters.

{\bf Acknowledgments} We would like to thank M. Muschol and A.
Minton for valuable discussions.  This work was supported by KSU
startup funds.

\appendix

\section{Mean Spherical Approximation MSA}

The attractive term $f_{att}$ in Eq. \ref{eq:Ftotal} can be given by
the MSA \cite{Tang2003} for a Yukawa fluid. As a second-order type
theory, the MSA yield
\begin{align}
f_{\text{att}} & = -\frac{1}{2}\sum_{n=1}^{\infty} \frac{v_n}{n}(\beta \epsilon)^{n} \notag \\
&= \frac{\alpha_0}{\Phi_0} \beta \epsilon - \frac{z^3}{6 \eta} \sum_{n=2}^{\infty}\bigg(\frac{2^{n-2} - n + 2}{n}\bigg) X^n\notag \\
& + \frac{z^3}{6 \eta} \sum_{n=2}^{\infty}\bigg(\frac{2^{n-2} - n +
2}{n}\bigg) \bigg(1 + \frac{Y}{z \psi}\frac{d}{d Y}\bigg){Y}^n
\label{eqatt}
\end{align}
where
\begin{equation}
\alpha_0 = \frac{L(z)}{z^2 (1-\eta)^2}; \label{eqA23}
\end{equation}
\begin{equation}
L(z) = 12 \eta [(1+\eta/2)z + 1 + 2 \eta] \label{eqA24}
\end{equation}

\begin{equation}
S(z) = (1-\eta)^2 z^3 + 6\eta(1-\eta)z^3 + 18 \eta^2z -12 \eta(1+2\eta) \label{eqA25}
\end{equation}
\begin{equation}
\Phi_0 = \frac{\exp({-z}) L(z) + S(z)}{z^3 (1-\eta)^3};
\label{eqA25}
\end{equation}
\begin{equation}
\psi = z^2(1-\eta)^2\frac{1-\exp(-z)}{\exp(-z)L(z)+S(z)}-12\eta(1-\eta)\frac{1-z/2-(1+z/2)\exp(-z)}{\exp(-z)L(z)+S(z)}
\end{equation}
\begin{equation}
w = \frac{6 \eta}{\Phi_0^2} \label{eqA26}
\end{equation}
\begin{equation}
v_n = \frac{2(2^{n-2} - n +2) w^{n-1} [(1+z\psi)^n - nz^{n-1}\psi^{n-1} - z^n \psi^n]}{z^{2n-3} \Phi_0^2}
\label{eqA27}
\end{equation}
\begin{equation}
X = \frac{(1+ z \psi)w}{z^2}\beta \epsilon
\label{eqA28}
\end{equation}
and
\begin{equation}
Y = \bigg(\frac{ w \psi}{z}\bigg) \beta \epsilon
\label{eqA29}
\end{equation}
By doing some algebric manipulation the free energy takes the simple
form
\begin{equation}
f_{att}=\frac{\alpha_0}{\Phi_0} \beta \epsilon - \frac{z^3}{6 \eta}
\bigg[\mathcal{F}(X) - \mathcal{F}(Y) - (X - Y) \frac{d
\mathcal{F}(Y)}{d Y}\bigg] \label{eqA30}
\end{equation}
where the function $\mathcal{F}(X) $
\begin{equation}
\mathcal{F}(X) = -\frac{1}{4} \ln(1-2X) - 2 \ln(1-X) - \frac{3}{2}X - \frac{1}{(1-X)} +1
\label{eqA31}
\end{equation}
and its first derivative is
\begin{equation}
\frac{d \mathcal{F}(X)}{d X} = \frac{X (1 - 3X + 3X^2)}{(1 - 2X)(1-X)^2}.
\label{eqA32}
\end{equation}

\section{Linearized Potential}

Here we solve for the ion distributions using the cell
approximation.  We model the protein as a charged sphere embedded in
a spherical cavity of solvent with radius $b=a \eta^{-1/3}$.  To do
this we need to solve the Poisson-Boltzmann equation
\begin{equation}
\epsilon \nabla_r^2 \Psi(\vec{r})=-\rho_f(\vec{r})-e N_A(c_0 e^{-e
\Psi(\vec{r})/k_BT}-c_0 e^{e \Psi(\vec{r})/k_BT}).
\label{eq:PBappdx}
\end{equation}
for the electrostatic potential $\Psi$.  Here $\rho_f$ is the fixed
charge distribution on the protein, $c_0$ is the bulk salt
concentration, $N_A$ is Avogadro's constant, and $\epsilon \simeq
80\epsilon_0$ is permeability of water. First, we put the PB
equation in dimensionless form
\begin{equation}
\nabla_y^2 \Phi=\sinh (\Phi)
\end{equation}
where $\Phi=e_p\Psi/k_BT$ is the dimensionless potential, $y=\kappa
r$ is a dimensionless length, and
\begin{equation}
\kappa^2=\frac{2 e_p^2 N_A c_0}{\epsilon_0\epsilon_w k_BT}.
\end{equation}
Now we linearize the potential around the local potential
$\Phi(y)=\phi(y)+\phi_0$, so we have
\begin{eqnarray}
\nabla_y^2 \Phi &=& \sinh (\phi+\phi_0) \\
&\simeq& \phi \cosh \phi_0 + \sinh \phi_0 \\
\nabla_x^2 \phi &=& \phi + \tanh \phi_0, \label{eq:PBx}
\end{eqnarray}
where the new length variable is
\begin{eqnarray}
x&=&\sqrt{\cosh \phi_0}y \\
&=&\sqrt{\cosh \phi_0}\kappa r.
\end{eqnarray}
The solution to Eq. \ref{eq:PBx} is
\begin{equation}
\phi(x)=A\frac{e^{-x}}{x}+B\frac{e^{x}}{x}-\tanh \phi_0,
\end{equation}
where the constants $A$ and $B$ are determined by the boundary
conditions.  The boundary conditions are that at the inner sphere
$r=a$ the electric field is equal to that of the bare macroion, and
at the outer sphere $r=b$ the electric field vanishes (due to charge
neutrality).  After scaling the sphere radii
\begin{equation}
\alpha  =  \sqrt{\cosh \phi_0}\kappa a
\label{eq:alpha}
\end{equation}
\begin{equation}
\beta  =  \sqrt{\cosh \phi_0}\kappa b
\label{eq:beta}
\end{equation}
the inner boundary condition becomes
\begin{eqnarray}
-E_0&=&\frac{d\phi}{dx} \\
&=&-A\frac{e^{-\alpha}}{\alpha}\left(1+\frac{1}{\alpha}\right)+B\frac{e^{\alpha}}{\alpha}\left(1-\frac{1}{\alpha}\right),
\label{eq:BCa}
\end{eqnarray}
where the dimensionless electric field is
\begin{equation}
E_0=\frac{qe_p\kappa \sqrt{\cosh \phi_0}}{4 \pi \epsilon k_BT
\alpha^2}, \label{eq:E0}
\end{equation}
where $qe_p$ is the charge on the central sphere.  At the outer
sphere boundary we have
\begin{equation}
-A\frac{e^{-\beta}}{\beta}\left(1+\frac{1}{\beta}\right)+B\frac{e^{\beta}}{\beta}\left(1-\frac{1}{\beta}\right)=0,
\end{equation}
from which we derive
\begin{equation}
B=Ae^{-2\beta}\frac{1+1/\beta}{1-1/\beta}. \label{eq:BofA}
\end{equation}
Combining Eqs. \ref{eq:BCa} and \ref{eq:BofA} we find
\begin{equation}
A[-e^{-\alpha}(\alpha +1)(\beta -1)+e^{\alpha - 2\beta}(\alpha
-1)(\beta+1)]=-\alpha^2 E_0 (\beta -1).
\end{equation}
So the two constants are
\begin{eqnarray}
A&=&\frac{\alpha^2 E_0 e^\beta
(\beta-1)}{e^{-\alpha+\beta}(\alpha+1)(\beta-1)-e^{\alpha-\beta}(\alpha-1)(\beta+1)}
\\
B&=&\frac{\alpha^2 E_0 e^{-\beta}
(\beta+1)}{e^{-\alpha+\beta}(\alpha+1)(\beta-1)-e^{\alpha-\beta}(\alpha-1)(\beta+1)}.
\end{eqnarray}
The total potential is then
\begin{eqnarray}
\Phi(x)&=&\phi(x)+\phi_0 \\
&=&\frac{\alpha^2
E_0}{e^{-\alpha+\beta}(\alpha+1)(\beta-1)-e^{\alpha-\beta}(\alpha-1)(\beta+1)}\left(\frac{e^{\beta-x}(\beta-1)}{x}+\frac{e^{x-\beta}(\beta+1)}{x}\right)
\nonumber \\
&&-\tanh \phi_0 + \phi_0, \label{eq:shiftedDHpotappdx}
\end{eqnarray}
so the constant used in the text is
\begin{equation}
C=\frac{\alpha^2
E_0}{e^{-\alpha+\beta}(\alpha+1)(\beta-1)-e^{\alpha-\beta}(\alpha-1)(\beta+1)}.
\end{equation}

The coulomb energy is obtained as follows
\begin{eqnarray}
f_{\rm coulomb}&=&\frac{\epsilon}{2}\int_\alpha^\beta |\vec{E}(\vec{r})|^2 d^3r \\
&=&\int_\alpha^\beta\bigg[\frac{1}{x^2}(A_1^2e^{-2x}+2A_1B_1+B_1^2e^{2x}) \nonumber \\
&&-\frac{2}{x}(-A_1^2e^{-2x}+B_1^2e^{2x}) + (A_1^2e^{-2x}+B_1^2e^{2x}-2A_1B_1)\bigg]dx\\
&=&\left[-\frac{1}{x}(A_1e^{-x}+B_1e^{x})^2+\frac{1}{2}(B_1^2e^{2x}-A_1^2e^{-2x}-4A_1B_1x)\right]^\beta_\alpha
\\
&=&-2C^2\beta^3+\frac{C^2}{\alpha}((\beta-1)e^{-\alpha+\beta}+(\beta+1)e^{\alpha-\beta})^2
\nonumber \\
&&-\frac{C^2}{2}((\beta+1)^2e^{2(\alpha-\beta)}-(\beta-1)^2e^{-2(\alpha-\beta)}-4(\beta^2-1)\alpha)
\end{eqnarray}
where $A_1 = C(\beta-1) \exp(\beta)$ and $B_1 = C(\beta+1)
\exp(-\beta)$.


\begin{thebibliography}{10}

\bibitem{Zimmerman1993} Zimmerman, Steven B and Minton, Allen P,
Annual review of biophysics and biomolecular structure, 27, {\bf 22}, 1993

\bibitem{Shire2004} Shire, Steven J and Shahrokh, Zahra and Liu, Jun,
Journal of pharmaceutical sciences, 6, 1390, {\bf 93}, 2004

\bibitem{McPherson2004} Mcpherson, Alexander, Methods (San Diego, Calif.),
3, 254, {\bf 34}, 2004

\bibitem{Tavares:2004b} Tavares, Frederico W. and Bratko, D. and Blanch, Harvey W and Prausnitz, John M.,
J. Phys. Chem. B, 26, 9228, {\bf 108}, 2004

\bibitem{Elcock2001} Elcock, Adrian H and McCammon, J. Andrew,
Biophysical journal, 2, 613, {\bf 80}, 2001

\bibitem{Coen1996} Coen, Cj and Newman, J and Blanch, Harvey W Hw and Prausnitz, John M., Journal of colloid and interface science, 1,
 276, {\bf 177}, 1996

\bibitem{Schmit2011gel} Schmit, Jeremy David and Whitelam, Stephen and Dill, Ken A,
The Journal of Chemical Physics, 8, 085103, {\bf 135}, 2011

\bibitem{tenWolde1997} ten Wolde, P. R. and Frenkel, Daan, Science, 5334,
1975, {\bf 277}, 1997

\bibitem{Hagen1994} Hagen, M. H. J. and Frenkel, Daan, The Journal of Chemical Physics, 5, 4093, {\bf 101}, 1994

\bibitem{Lomakin2003} Lomakin, Aleksey and Asherie, Neer and Benedek, George B.,
Proceedings of the National Academy of Sciences of the United States of America,
 18, 10254, {\bf 100}, 2003

\bibitem{Lomakin1996} Lomakin, Aleksey and Asherie, Neer and Benedek, George B.,
The Journal of Chemical Physics,
4, 1646, {\bf 104}, 1996

\bibitem{Asherie:1996} Asherie, N and Lomakin, A and Benedek, George B.,
Phys. Rev. Lett., 23, 4832, {\bf 77}, 1996

\bibitem{Gogelein2008} G\"{o}gelein, Christoph and N\"{a}gele, Gerhard and Tuinier, Remco and Gibaud, Thomas and Stradner, Anna and Schurtenberger, Peter,
The Journal of chemical physics, 8, 085102, {\bf 129}, 2008

\bibitem{Fu2003} Fu, Dong and Li, Yigui and Wu, Jianzhong,
Physical Review E, 1, 38, {\bf 68}, 2003

\bibitem{Tang2003} Tang, Yiping, The Journal of Chemical Physics, 9,
4140, {\bf 118}, 2003


\bibitem{Durand-Vidal2000} Durand-Vidal, S. and Simonin, J.-P. and Turq, P.,
Springer, Electrolytes at Interfaces (Progress in Theoretical Chemistry and Physics), 2000

\bibitem{Duh1997} Duh, D.M. and Mier-Y-Teran, L, Molecular Physics,
3, 373,  {\bf 90}, 1997


\bibitem{Carnahan1969} Carnahan, N.F. and Starling, K.E., The Journal of Chemical Physics,
635, {\bf 51}, 1969

\bibitem{Wall1957} Wall, F. T. and Berkowitz, Joan, The Journal of Chemical Physics, 1,
114, {\bf 26}, 1957


\bibitem{Levin2003} Levin, Yan and Trizac, Emmanuel and Bocquet, Lyd\'{e}ric,
Journal of Physics: Condensed Matter, 48, S3523, {\bf 15}, 2003


\bibitem{Kanal1994} Kanal, K M and Fullerton, G D and Cameron, I L,
Biophysical journal,
1, 153, {\bf 66}, 1994


\bibitem{Vilker1981} Vilker, V.L. and Colton, C.K. and Smith, K.A., Journal of Colloid and Interface Science,
2, 548, {\bf 79}, 1981


\bibitem{Minton1982} Minton, Allen P. and Edelhoch, H., Biopolymers, 2,
451, {\bf 21}, 1982


\bibitem{Minton1995} Minton, Allen P., Biophysical chemistry, 1, 65,
{\bf 57}, 1995

\bibitem{Salis2011} Salis, Andrea and Bostr\"{o}m, Mathias and Medda, Luca and Cugia, Francesca and Barse, Brajesh and Parsons, Drew F and Ninham, Barry W and Monduzzi, Maura,
Langmuir : the ACS journal of surfaces and colloids,
18, 11597, {\bf 27}, 2011


\bibitem{Piazza:1999} Piazza, Roberto, J. Cryst. Growth, 2-4,
415, {\bf 196}, 1999


\bibitem{Chaikin2000} Chaikin, P. M. and Lubensky, T.C., Cambridge University Press,
Principles of Condensed Matter Physics, 2000


\bibitem{Muschol:1997} Muschol, Martin and Rosenberger, Franz,
J. Chem. Phys., 6, 1953, {\bf 107}, 1997


\bibitem{Kern:2003} Kern, Norbert and Frenkel, Daan, J. Chem. Phys.,
21, 9882, {\bf 118}, 2003


\bibitem{Bianchi:2008} Bianchi, Emanuela and Tartaglia, Piero and Zaccarelli, Emanuela and Sciortino, Francesco,
J Chem Phys, 14, 144504, {\bf 128}, 2008


\bibitem{Liu:2009} Liu, Hongjun and Kumar, Sanat K and Sciortino, Francesco and Evans, Glenn T,
J. Chem. Phys., 4, 044902, {\bf 130}, 2009


\bibitem{Wang2011a}, Wang, Ying and Lomakin, Aleksey and Latypov, Ramil F and Benedek, George B.,
Proceedings of the National Academy of Sciences of the United States of America,
40, 16606, {\bf 108}, 2011


\bibitem{Muschol1995}
Muschol, Martin and Rosenberger, Franz, J. Chem. Phys.,
24, 10424, {\bf 103}, 1995


\end{thebibliography}

\end{document}